\documentclass[12pt,a4paper]{article}
\usepackage{graphicx}
\usepackage[caption=false]{subfig}
\usepackage{epsfig}
\usepackage{amssymb}
\usepackage{amsmath}
\usepackage{lipsum}
\usepackage{subfig}
\usepackage[left=2cm,right=2cm,top=2cm,bottom=2cm]{geometry}
\usepackage{multirow}
\usepackage{amsmath}
\usepackage[toc]{appendix}
\usepackage{color,soul}
\begin{document}
\title{Charge radii of heavy flavored mesons in a potential model}
\author{Tapashi Das$^{1,\ast}$, D K Choudhury$^{1,2}$, K K Pathak$^{3}$ and N S Bordoloi$^{4}$
\\  $^1$Department of Physics,\ Gauhati University,
Guwahati-781014, India.\\
$^2$ Physics Academy of North-East, Guwahati-781014, India. \\
$^3$ Department of Physics, Arya Vidyapeeth College, Guwahati-781016, India\\
$^4$ Department of Physics, Cotton College, Guwahati-781001, India\\
$^\ast$\small{email:t4tapashi@gmail.com}}
\date{}
\maketitle

\begin{abstract} 
We report the results for charge radii of heavy flavored mesons ($D^+, D^0, D_s^+, B^+, B^0, B_s^0$) in an improved QCD potential model. To enhance the effectiveness of short range and long range effect of the potential $V(r)=-\frac{4 \alpha_s}{3r}+br$ in the perturbative procedure a cut-off parameter $r^P$ is introduced as an integration limit. The obtained results are found to be comparable with other available data. The limitation of the approach is discussed in the manuscript.\\
\end{abstract}



\section{Introduction}
For both qualitative and quantitative descriptions of hadron spectroscopy, the potential model description in the non-relativistic regime of QCD \cite{qcd} is found to be tremendously successful. \\
 
In our work, we consider Coulomb plus linear Cornell potential \cite{vipt}, 

\begin{equation}
V(r)=-\frac{4\alpha_s}{3r}+br+c.
\end{equation}

This potential is very useful to apply in the quantum mechanical perturbation theory in the study of heavy flavored mesons, since at short distance, linear term is effectively considered as perturbation and at long distance Coulomb potential is considered as perturbation. i.e. it is based on the two kinds of asymptotic behaviors: ultraviolet at short distance (Coulomb like) and infrared at large distance (linear confinement term). In the potential (1), -$\frac{4}{3}$ is due to the color factor, $\alpha_s$ is the strong coupling constant, $r$ is the inter-quark distance, $b$ is the confinement parameter and `$c$' is a constant scale factor which is a phenomenological constant and is needed to reproduce correct masses of heavy-light meson bound state.\\

In the present work we have considered the scaling factor $c=0$ as done in ref. \cite{my1, my2, my3}. Because in general, it is expected that a constant term `$c$' in the potential should not affect the wave function of the system while applying the perturbation theory. But in ref. \cite{my2} it is seen that whether the term `$c$' is in parent or perturbed part of the Hamiltonian, it always appears in the total wave function which is inconsistent with the quantum mechanical idea that a constant term `$c$' in the potential can at best shift the energy scale, but should not perturb the wave function i.e. a Hamiltonian $H$ with such a constant and another $H^\prime$ without it should give rise to the same wave functions.\\

Here we introduce a cut-off parameter  $r^P$ as an integration limit and compute the charge radii of $D^+, D^0, D_s^+, B^+, B^0, B_s^0$ mesons. It is well known that at short distance Coulomb potential plays a more dominant role than the linear confinement of potential (1) and at large distance the confinement takes over the Coulomb effect. Therefore if the inter-quark separation `$r$' can be roughly divided into two regions $0<r<r^P$ for short distance and $r^P<r<r_0$ for long distance effectively, where $r^P$ is the point where one of the potentials will dominate over the other. In such situation confinement parameter ($b$) and the strong coupling parameter ($\alpha_s$) can be considered as effective and appropriate small pertubative parameters. Therefore in the present work of the paper we tried to incorporate both the short range and long range effect of the potential in the construction of total wave function.\\

We use the two-body Schrodinger's equation and obtain the first-order perturbed wave function of the QCD potential using Dalgarno's method \cite{dalgarno1, dalgarno2} of perturbation.\\

In this paper, the obtained results for charge radii of mesons are compared with earlier work \cite{nsbijp,bjh, nsbnew} and also with the prediction of other model values \cite{hwang, fernandez, aitchison,  lombard}. Limitation of the approach is also discussed.\\

The paper is organised as follows: in section 2, we outline the formalism where we have discussed the model and obtain the expressions for form factors. In section 3, we summarize the results and discussions. Section 4 contains conclusion.

\section{Formalism}
\subsection{The Model}
The non-relativistic two body Schrodinger equation \cite{se} is
\begin{equation}
H|\psi\rangle=(H_0+H^\prime)|\psi\rangle=E|\psi\rangle,
\end{equation}

so that the first-order perturbed eigenfunction $\psi^{(1)}$ and eigenenergy $W^{(1)}$ can be obtained using the relation
\begin{equation}	
H_0 \psi^{(1)} + H^\prime \psi^{(0)}=W^{(0)}\psi^{(1)} + W^{(1)} \psi^{(0)},
\end{equation}

where
\begin{equation}
W^{(0)}=  <\psi^{(0)}\vert H_0 \vert \psi^{(0)}>,
\end{equation}
\begin{equation}
W^{(1)}= <\psi^{(0)}\vert H^\prime \vert \psi^{(0)}>.
\end{equation}

We calculate the total wave function using Dalgarno's method of perturbation for the potential (1) with $c=0$,\\

\begin{equation}
V(r)=-\frac{4\alpha_s}{3r}+br.
\end{equation} 

The two choices for parent and perturbed Hamiltonian are\\

choice-I: $H_0=-\frac{\nabla^2}{2\mu}-\frac{4\alpha_s}{3r}$ as parent and $H^\prime=br$ as perturbation and\\

choice-II: $H_0=-\frac{\nabla^2}{2\mu}+br$ as parent and $H^\prime=-\frac{4\alpha_s}{3r}$ as perturbation.\\

From choice-I and II, we can find the bounds on $r$ upto which both the choices are valid. \\

From choice-I,
\begin{equation}
\label{rs}
\mid -\frac{4\alpha_s}{3r}\mid \textgreater \mid br \mid
\end{equation}

and from choice-II,
\begin{equation}
\label{rl}
\mid br \mid \textgreater \mid -\frac{4\alpha_s}{3r}\mid.
\end{equation}

Inequality (\ref{rs}) and (\ref{rl}) will correspond to a particular point $r$, say $r^P$, where $r^P=\sqrt{\frac{4\alpha_s}{3b}}$ such that for the short distance, i.e. $r<r^P$ Coulomb part is dominant over the linear confinement term and for long distance, i.e. $r>r^P$ linear part is dominant over the Coulombic term. Thus the point $r^P$ measures the distance at which the potential changes from being dominantly Coulombic $(r<r^P)$ to dominantly linear $(r>r^P)$. At potential level, the continuity at a particular point of $r$ is quite clear as is evident from Fig.1 of ref. \cite{vipt}.\\

The total wave function for choice-I is 
\begin{equation}
\label{coulomb}
\psi_I^{total}(r)=\frac{N}{\sqrt{\pi a_0^3}}\left[ 1-\frac{1}{2}\mu b a_0r^2\right]\left( \frac{r}{a_0}\right) ^{-\epsilon} e^{-\frac{r}{a_0}},
\end{equation}

where the normalization constant $N$ is

\begin{equation}
\label{psiCN}
N=\left[ \int_0^{r^P} \frac{4 r^2}{a_0^3}\left[ 1-\frac{1}{2}\mu b a_0r^2\right]^2\left( \frac{r}{a_0}\right) ^{-2\epsilon} e^{-\frac{2r}{a_0}}dr\right] ^{-\frac{1}{2}}
\end{equation}

and

\begin{equation}
a_0=\left( \frac{4}{3}\mu\alpha_s\right)^{-1},
\end{equation}

\begin{equation}
\mu=\frac{m_i m_j}{m_i+m_j},
\end{equation}

$m_i$ and $m_j$ are the masses of the quark and antiquark respectively, $\mu$ is the reduced mass of the mesons and

\begin{equation}
\epsilon=1-\sqrt{1-\left( \frac{4}{3}\alpha_s\right) ^2}
\end{equation}

is the correction for relativistic effect \cite{re1, re2}due to Dirac modification factor.\\

Similarly, the total wave function upto $O(r^4)$ for choice-II is

\begin{equation}
\label{linear}
\psi_{II}^{total}(r)=\frac{N^\prime}{r} \left[1+A_0r^0+A_1(r)r+A_2(r)r^2+A_3(r)r^3+A_4(r)r^4\right]A_i[\rho_1 r+\rho_0] \left( \frac{r}{a_0}\right) ^{-\epsilon},
\end{equation}

where $A_i[r]$ is the Airy function \cite{airy} and $N^\prime$ is the normalization constant,

\begin{equation}\footnotesize
\label{psiLN}
N^\prime= \left[ \int_{r^P}^{r_0} 4 \pi \left[1+A_0r^0+A_1(r)r+A_2(r)r^2+A_3(r)r^3+A_4(r)r^4\right]^2 \left( A_i[\rho_1 r+\rho_0]\right) ^2 \left( \frac{r}{a_0}\right) ^{-2\epsilon}dr\right]^{-\frac{1}{2}}.
\end{equation}

Even though the Airy's function vanishes exponentially as $r\rightarrow \infty$ \cite{airy} and is normalizable too, the additional cut-off $r_0$ is used in the integration basically due to the polynomial approximation of the series expansion used in the Dalgarno's method of perturbation and is independent of the property of the Airy function.\\

The co-efficients $A_0$, $A_1$, $A_2$, $A_3$ and $A_4$ of the series solution as occured in Dalgarno's method of perturbation, are the function of $\alpha_s, \mu$ and $b$:
\begin{equation}
A_0=0,
\end{equation}

\begin{equation}
A_1=\frac{-2\mu \frac{4\alpha_s}{3}}{2\rho_1 k_1+\rho_1^2 k_2},
\end{equation}

\begin{equation}
A_2=\frac{-2\mu W^1}{2+4 \rho_1 k_1+ \rho_1^2 k_2},
\end{equation}

\begin{equation}
A_3=\frac{-2\mu W^0 A_1}{6+6 \rho_1 k_1+ \rho_1^2 k_2},
\end{equation}

\begin{equation}
A_4=\frac{-2\mu W^0 A_2+2\mu b A_1}{12+8 \rho_1 k_1+ \rho_1^2 k_2}.
\end{equation}

The parameters:
\begin{equation}
\rho_1=(2\mu b)^{\frac{1}{3}},
\end{equation}

\begin{equation}
\rho_0=-\left[ \frac{3\pi (4n-1)}{8}\right] ^{\frac{2}{3}}
\end{equation}

(in our case n=1 for ground state),

\begin{equation}
k=\frac{0.355-(0.258) \rho_0}{(0.258) \rho_1},
\end{equation}

\begin{equation}
k_1=1+\frac{k}{r},
\end{equation}

\begin{equation}
k_2=\frac{k^2}{r^2},
\end{equation}

\begin{equation}
W^1=\int \psi^{(0)\star} H^{\prime} \psi^{(0)} d\tau,
\end{equation}

\begin{equation}
W^0=\int \psi^{(0)\star} H_0 \psi^{(0)} d\tau.
\end{equation}

\subsection{Form factor and charge radii}

The elastic charge form factor for a charged system of point quarks has the $Q^2$ dependent form \cite{ff1}

\begin{equation}
\label{ff}
F(Q^2)=\sum_{i=1}^2 \frac{e_i}{Q_i}\int_0^\infty 4\pi r \mid\psi(r)\mid^2 sin(Q_ir)dr,
\end{equation}

where $Q^2$ is the four momentum transfer square and $e_i$ is the charge of the $i^{th}$ quark/antiquark and 

\begin{equation}
Q_i=\frac{\sum_{j\neq i} m_jQ}{\sum_{i=1}^2 m_i},
\end{equation}

where $Q_i$ describes how the virtuality $Q^2$ is shared between the quark and antiquark pair of the meson.\\

In the present model, we redefine equation (\ref{ff}) as

\begin{equation}
\label{ff1}
\begin{split}
F(Q^2)\approx \sum_{i=1}^2 \frac{e_i}{Q_i}\int_0^{r^P} 4\pi r \mid\psi_I(r)\mid^2 sin(Q_ir)dr \\
+ \sum_{i=1}^2 \frac{e_i}{Q_i}\int_{r^P}^{r_0} 4\pi r \mid\psi_{II}(r)\mid^2 sin(Q_ir)dr\\
\end{split}
\end{equation}

\begin{equation}
\label{ff2}
F(Q^2)\approx F(Q^2)\mid_I + F(Q^2)\mid_{II}.
\end{equation}

In equations (\ref{ff1}) and (\ref{ff2}), we have used approximation signs rather than equals because $F(Q^2)\mid_I$ will give approximate results for each form factor when $r\le r^P$ and $F(Q^2)\mid_{II}$ will give approximate results when $r\ge r^P$.\\
%To obtain the results we have used Mathematica version 7.0.0 where we have considered Airy function upto $O(r^4)$. 

To check the behavior of the form factor with momentum transfer square $Q^2$ we obtain the analytic expressions for form factors considering Airy's function upto order $r^1$ as shown in Appendix A and B.\\

\textbf{(a) With Dirac modification factor:}\\

The $1^{st}$ part of (\ref{ff2}), $F(Q^2)\mid_I$ is solved using Coulomb potential as parent and linear potential as perturbation wave function (\ref{coulomb}) with relativistic correction (as shown in Appendix A) which gives 

\begin{equation}
\label{ff3}
\begin{split}
F(Q^2)\mid _I\approx N^2 \sum_{i=1}^2 e_i \biggl[ \frac{1}{2^{1-2\epsilon}}\gamma(2-2\epsilon, r^P)(2-2\epsilon)\frac{1}{(1+\frac{a_0^2 Q_i^2}{4})^{\frac{3}{2}-\epsilon}} \\ 
- \frac{\mu b a_0^3}{2^{3-2\epsilon}} \gamma(4-2\epsilon, r^P)(4-2\epsilon)\frac{1}{(1+\frac{a_0^2 Q_i^2}{4})^{\frac{5}{2}-\epsilon}}\\ 
+\frac{\mu^2 b^2 a_0^6}{2^{7-2\epsilon}} \gamma(6-2\epsilon, r^P)(6-2\epsilon)\frac{1}{(1+\frac{a_0^2 Q_i^2}{4})^{\frac{7}{2}-\epsilon}}\biggr],
\end{split}
\end{equation}

where the Incomplete Gamma function $\gamma(s,r^P)$ is defined as

\begin{equation}
\int_0^{r^P} t^{s-1} e^{-t}dt= \gamma(s,r^P).
\end{equation}

From the reality condition of equation (\ref{ff3}), we get that $0<\epsilon<1$, thus form factor falls with the increasing value of $Q^2$.\\

Similarly, the $2^{nd}$ part of integration (\ref{ff2}), $F(Q^2)\mid_{II}$ is solved using wave function (\ref{linear}) (as shown in Appendix B), which gives

\begin{equation}
\label{ff4}
\begin{split}
F(Q^2)\mid _{II}\approx 4 \pi N^{\prime 2} a_0^{2 \epsilon} \sum_{i=1}^2 e_i \biggl[
\sum_{k=1}^{11} F_k\frac{1}{(Q_i^2)^{\frac{k-2\epsilon}{2}}}
\biggr],
\end{split}
\end{equation}

where $F_k$'s are defined in equation (B.5) of Appendix-B.\\

The constraint on equation (\ref{ff4}) is that for the term with $k=1$, $\epsilon<1$. \\

\textbf{(b) Without Dirac modification factor:}\\

The $1^{st}$ part of the integration (\ref{ff2}) $F(Q^2)\mid_I$ with $\epsilon=0$ gives\\

\begin{equation}
\begin{split}
F(Q^2)\mid _I\approx N^2 \sum_{i=1}^2 e_i \biggl[\gamma(2, r^P)\frac{1}{(1+\frac{a_0^2 Q_i^2}{4})^{\frac{3}{2}}}
-\frac{\mu b a_0^3}{2} \gamma(4, r^P)\frac{1}{(1+\frac{a_0^2 Q_i^2}{4})^{\frac{5}{2}}} \\ 
+\frac{3\mu^2 b^2 a_0^6}{2^6} \gamma(6, r^P)\frac{1}{(1+\frac{a_0^2 Q_i^2}{4})^{\frac{7}{2}}}
\biggr].
\end{split}
\end{equation}

Similarly, with $\epsilon=0$, $F(Q^2)\mid_{II}$ is

\begin{equation}
F(Q^2)\mid _{II}\approx 4 \pi N^{\prime 2}\sum_{i=1}^2 e_i \biggl[
\sum_{k=2}^{11} F^\prime_k\frac{1}{(Q_i^2)^{\frac{k}{2}}}\biggr],
\end{equation}

where $F^\prime_k$'s are defined in equation (B.8) of Appendix-B.\\

The average charge radii square for the mesons is extracted from the form factors at their low $Q^2$ behavior using the relation \cite{hwang},

\begin{equation}
\label{charge radii}
\langle r^2 \rangle= -6 \frac{d^2}{dQ^2}F(Q^2)\vert_{Q^2=0}\approx -6 \frac{d^2}{dQ^2}\left[ F(Q^2)\mid _{I}+F(Q^2)\mid _{II}\right] \vert_{Q^2=0}.
\end{equation}

%\begin{equation}
%\approx -6 \frac{d^2}{dQ^2}\left[ F(Q^2)\mid _{I}+F(Q^2)\mid _{II}\right] \vert_{Q^2=0}
%\end{equation}
%With running coupling constant $\alpha_s$ and the number of quark flavors  $N_f$,

%\begin{equation}
%\alpha_s(Q^2)=\frac{12 \pi}{(33-2N_f)ln(\frac{Q^2}{\Lambda^2})}.
%\end{equation}

%For $100 \leq \Lambda \leq 300 MeV$ and for $3 \leq N_f \leq 5$, $\alpha_s$ is always around 0.2 when $Q=10 GeV$ and it varies slowly from $Q=5$ to $20 GeV$ \cite{qvalue}. Equation (33) says, the coupling constant decreases at large $Q^2$.

\section{Results and Discussions}

In Table \ref{rp}, we have recorded the numerical values of the cut-off parameter $r^P$ in $Fermi$ at charmonium and bottomonium scale.

\begin{table}[h]
\centering
\caption{$r^P$ in $Fermi$ with $c=0$ and $b=0.183 GeV^2$}
\begin{tabular}{ll}
\hline
            $\alpha_s$-value           & $r^P$ \\
            &($Fermi$)\\ \hline
                 0.39 &  0.332\\     
            (for charmonium scale)    &\\ \hline
           
                    0.22 & 0.249\\
             (for bottomonium scale)         &    \\
  \hline
\end{tabular}
\label{rp}
\end{table}             

\newpage
In Figure 1 we display the variation of form factor $F(Q^2)$ vs $Q^2$ for charged mesons ($D^+(c\bar{d})$, $D^+(c\bar{s})$ and $B^+(u\bar{b})$)and in Figure 2 we display the variation of form factor for neutral mesons ($D^0(c\bar{u})$, $B^0(d\bar{b})$ and $B^0_s(s\bar{b})$) respectively, using equation (\ref{ff1}) with Dirac modification factor. \\

The input parameters used are $m_u=0.336 GeV$, $m_d=0.336 GeV$, $m_s=0.483 GeV$, $m_c=1.55 GeV$, $m_b=4.95 GeV$ and $b=0.183 GeV^2$ and $\alpha_s$ values 0.39 and 0.22 for charmonium and bottomonium scale respectively and are same with the previous work \cite{my2, my3}. For our calculations, we set the cut-off ($r_0$) in the range of 1 $Fermi$ (5.076 $GeV^{-1}$) \cite{bali} for the wave function $\psi_{II}(r)$ .\\

\begin{figure}[!h]
\begin{center}
\subfloat[]{%
  \includegraphics[width=0.45\linewidth]{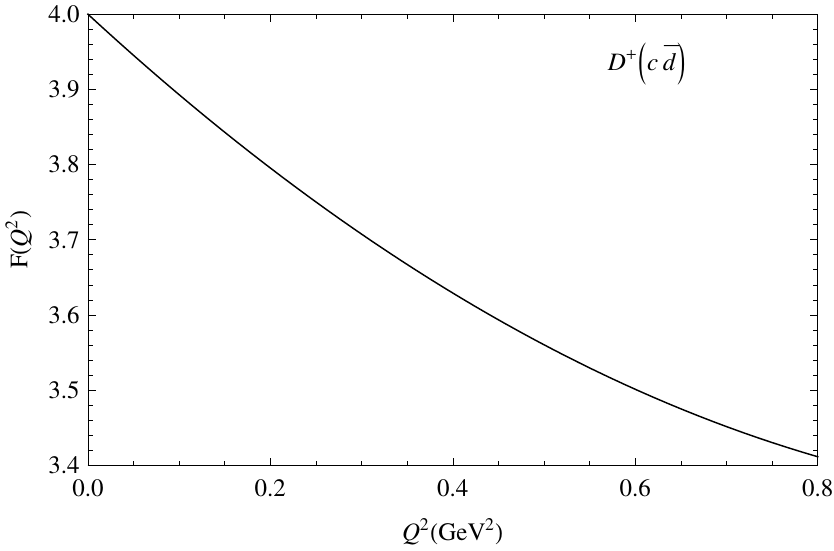}%
}\vspace{1ex}
\subfloat[]{%
  \includegraphics[width=0.45\linewidth]{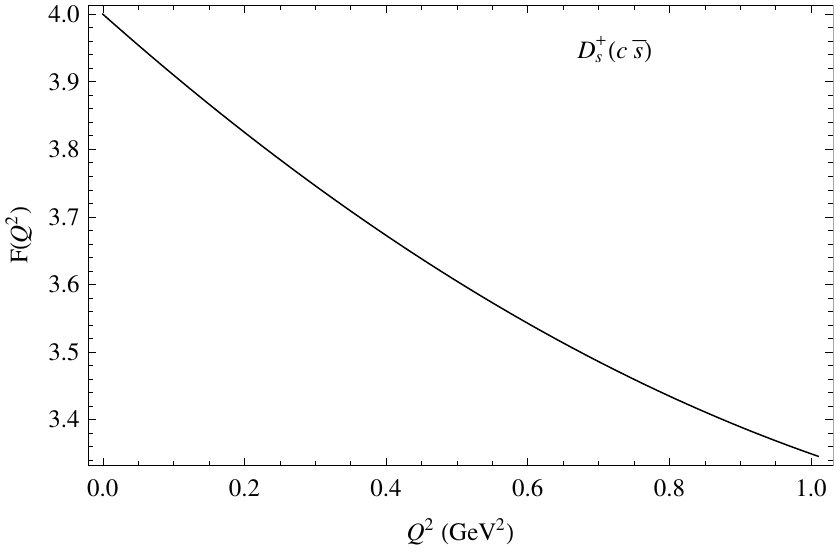}%
}\vspace{1ex}
\subfloat[]{%
  \includegraphics[width=0.45\linewidth]{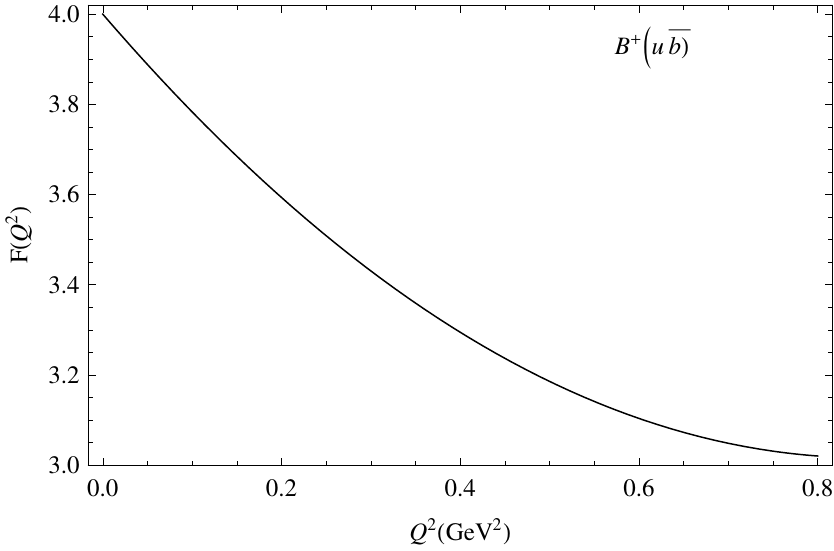}%
}\vspace{1ex}
\caption{Variation of form factor $F(Q^2)$ with $Q^2$ for a) $D^+(c\bar{d})$ meson, b) $D^+(c\bar{s})$ meson and c) $B^+(u\bar{b})$ meson.}
\label{fig:1}
\end{center}
\end{figure}

\begin{figure}[!h]
\begin{center}
\subfloat[]{%
 \includegraphics[width=0.45\linewidth]{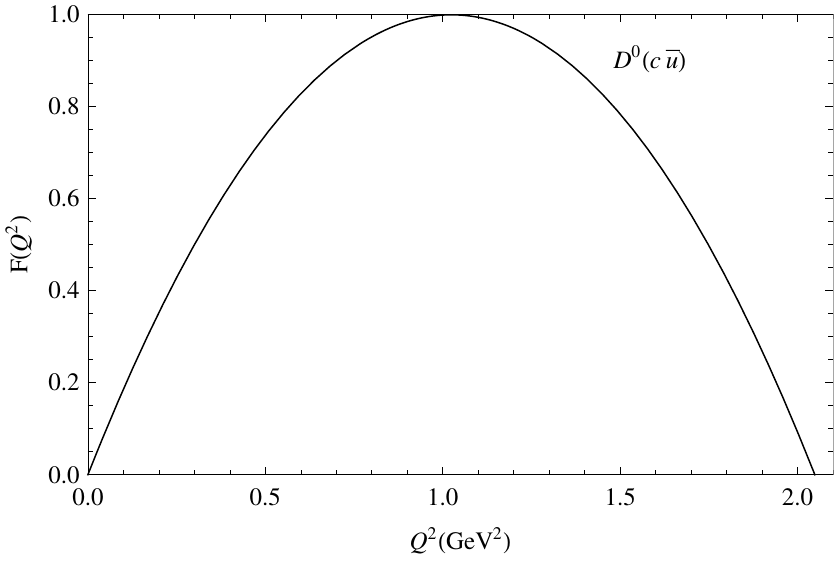}%
}\vspace{1ex}
\subfloat[]{%
  \includegraphics[width=0.45\linewidth]{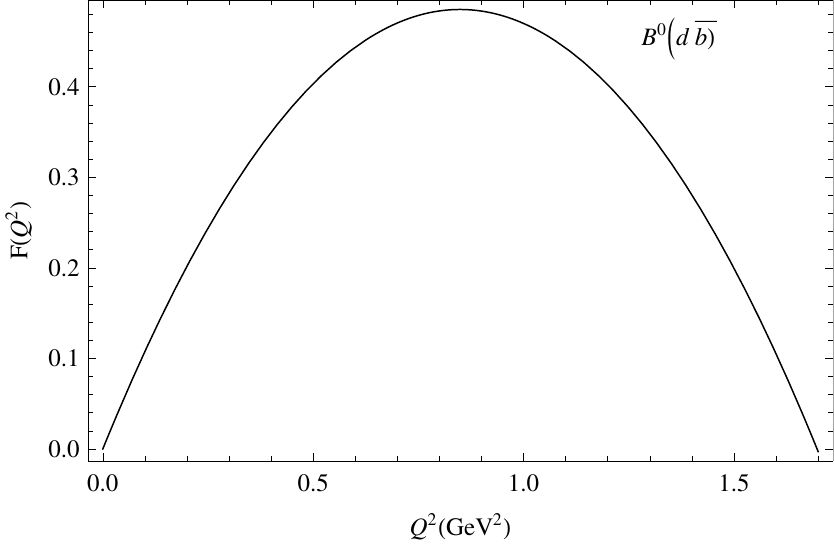}%
}\vspace{1ex}
\subfloat[]{%
  \includegraphics[width=0.45\linewidth]{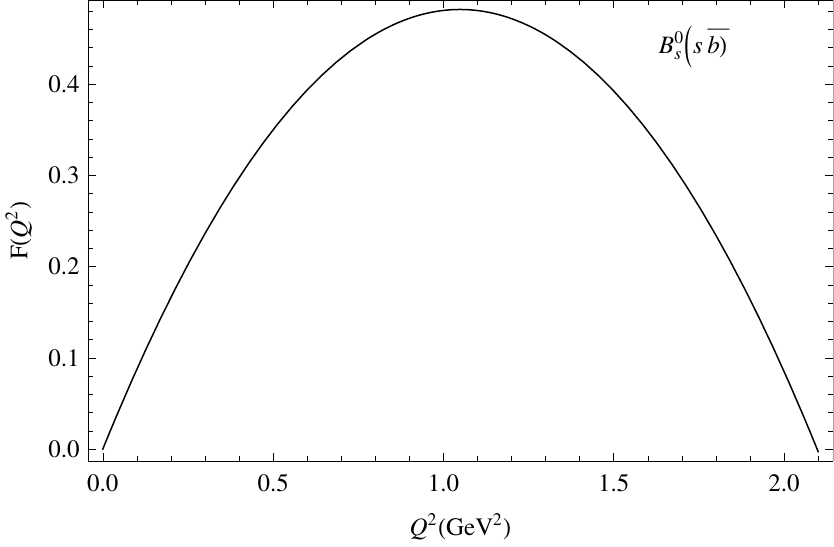}%
}\vspace{1ex}
\caption{Variation of form factor $F(Q^2)$ with $Q^2$ for a) $D^0(c\bar{u})$,  b) $B^0(d\bar{b})$ and c) $B^0_s(s\bar{b})$ meson.}
\label{fig2}
\end{center}
\end{figure}

\newpage
It seen that the form factor of the charged mesons (Figure 1) decreases with the increasing value of $Q^2$, but for neutral mesons (Figure 2) form factor first increases for small $Q^2$ and then decreases with the increasing value of momentum transfer square. A similar behavior for neutral pseudoscalar Kaon is also suggested in ref.\cite{kaon}. Our study also show a temporary rise in form factor does exist for heavy flavored neutral mesons near $Q^2 \approx 1 GeV^2$ (Figure 2). This is presumably due to the non-dominant behavior of small $Q^2$ over the other parameters involved. From the graphs, it is evident that the range of validity of the model is not beyond $\sim 2.1 GeV^2$.\\

In Table 2, we present the obtained results for the charge radii for various $D$ and $B$ mesons using equation (\ref{charge radii}) and compare them with the results of ref. \cite{nsbijp,bjh} and with the prediction of other models \cite{hwang,lombard}. 

\begin{table}[!h]\footnotesize
\caption{ The mean square charge radii of $D$ and $B$ mesons. }
\centering
\begin{tabular}{lllllll}
\hline
\multirow{3}{*}{Meson} &  \multicolumn{6}{c}{ $\langle r^2 \rangle$ ($Fermi^2$)} \\ \cline{2-7}
                       & \multicolumn{2}{c}{Present work} & \multirow{2}{*}{Previous} & \multirow{2}{*}{Previous} & \multirow{2}{*}{\cite{hwang}}  & \multirow{2}{*}{\cite{lombard}}\\ \cline{2-3}\\
                   & with  & without & work\cite{nsbijp} & work \cite{bjh}& & \\ 
                   &Dirac  & Dirac & & & & \\ 
                   & modification & modification & & & &\\ 
                    & factor & factor & & & &\\ \hline
 $D^+(c\bar{d}$)   &     0.260   & 0.265 &  0.134 & 0.011   &  0.184    & 0.219\\
 $D^0(c\bar{u})$   &    -0.453  & -0.463 & -0.234 & -0.013  &  -0.304  & -0.403\\  
 $D^+_s(c\bar{s})$ &     0.216   & 0.222 &  0.126 & 0.010   &   0.124   &-\\ 
 $B^+(u\bar{b})$   &     0.536  & 0.538 &  2.96  & 0.060   &   0.378   &-\\  
 $B^0(d\bar{b})$   &    -0.266   &-0.267 &  -1.47 & -0.030  &  -0.187   &-\\ 
 $B^0_s(s\bar{b})$ &    -0.214  & -0.215  &  -1.37 & -0.025  &  -0.119  &-\\    \hline
\end{tabular}
\label{charge radii table}
\end{table}
\newpage
From Table \ref{charge radii table} we can see that our predicted results for $D^+(c\bar{d})$ and $D^0(c\bar{u})$ mesons are in good agreement with those of ref. \cite{lombard}. Also the introduction of the Dirac modification factor doesn't change the results significantly. Further the present results for $B$ mesons are found to be improved than earlier analysis of ref. \cite{bjh}. In ref. \cite{bjh}, the charge radii of various heavy and light mesons were found to be very small where Variationally Improved Perturbation Theory (VIPT) \cite{vipt} was used. On the other hand in ref.\cite{nsbijp}, the charge radii of various mesons were calculated considering $b=0$, where the results for heavy flavored $D$ mesons were found to agree well with the experimental values but for heavy flavored $B$ mesons, the results were large compared to other theoretical models. The present results for $B$ mesons are found to be very much improved than the earlier analysis of ref.\cite{nsbijp,bjh}. In ref. \cite{nsbnew}, the charge radii of mesons were found for a non-zero value of scaling factor `$c$' and with $b=0.183 GeV^2$, where two-loop V-scheme was used for a large value of coupling constant $\alpha_v=0.625$.\\ 

From Table \ref{charge radii table} it is interesting to see that for all the neutral mesons the mean square charge radius is negative. A well explanation for negative charge radius square of the neutral meson can be found in ref. \cite{book charge}. Here let us explain this for neutral $D^0(c\bar{u})$ meson.\\

We define a center of mass coordinate for the quark antiquark $(Q\bar{q})$ bound state of meson,

\begin{equation}
\label{CM}
R=\frac{m_Q r_Q+m_{\bar{q}}r_{\bar{q}}}{m_Q +m_{\bar{q}}},
\end{equation}

where $r_Q$ and $r_{\bar{q}}$ are the heavy $(Q)$ and light anti-quark $(\bar{q})$ coordinates respectively.\\

The mean square charge radius of the meson can be written as the deviation from the center of mass coordinate squared weighted by the quark and antiquark constituents of the meson, which has the simplified form,

%\begin{equation}
%\label{weight}
%\langle r^2 \rangle=\langle\sum_{i=1}^2 e_i^2(r_i-R)^2\rangle
%\end{equation}

%\begin{equation}
%=\alpha \langle\sum_{i=1}^2 Q_i(r_i-R)^2\rangle,
%\end{equation}

%\hl{where $\alpha$ is....... and $Q_i$ is the charge of the quark/antiquark.}\\
\begin{equation}
\label{rD}
\langle r^2 \rangle_{D^0}= \frac{(Q_Q m_{\bar{q}}^2+Q_{\bar{q}}m_Q^2) \langle\delta^2 \rangle_{D^0}}{(m_Q+m_{\bar{q}})^2},
\end{equation}

where $Q_Q$ and $Q_{\bar{q}}$ are charge of the quark and anti-quark respectively.\\

$\delta=r_Q-r_{\bar{q}}$ is the relative coordinate.\\

For $D^0(c\bar{u})$ meson, $m_{\bar{q}}=m_u=m=0.336 GeV$\\

and $m_Q=m_c=1.55 GeV=\gamma m$; $\gamma=4.61$.\\

Thus from equation (\ref{rD}),
\begin{equation}
\label{RD}
\langle r^2 \rangle_{D^0}= \frac{2-2\gamma^2}{3+3\gamma^2}\langle\delta^2 \rangle_{D^0}.
\end{equation}

Since $\gamma=4.61$ and $\langle\delta^2 \rangle_{D^0}>0$, from equation (\ref{RD}), it is clear that $D^0(c\bar{u})$ has a negative square charge radius.\\

In $D^0(c\bar{u})$ meson, a negatively charged light $u$-antiquark is orbiting around a heavier positively charged $c$-quark. Since the mass of $c$-quark is very large compared to the $u$-antiquark, when we probe lightly into the charge distribution, we will see the charge of the light objects which are in the tail of the distribution orbiting out at large distances.\\

The same explanation is valid for $B^0(d\bar{b})$ and $B^0_s(s\bar{b})$ mesons also, where a light $d$-quark is orbiting around a heavier $b$-antiquark and a light $s$-quark is orbiting around a heavier $b$-antiquark respectively.\\

The perturbative stability of our results is also checked in the present model as shown in Table \ref{stability charge}.

\begin{table}[!h]
\caption{Mean square charge radii of $D$ and $B$ mesons. }
\centering
\begin{tabular}{lll}
\hline
                               Meson              &  \multicolumn{2}{c}{$\langle r^2 \rangle$ ($Fermi^2$)}   \\ \hline
                   & With Parent  & With Total   \\ 
                   & wave function &wave function \\ \hline
 $D^+(c\bar{d}$)   & 0.233&    0.260      \\ 
 $D^0(c\bar{u})$   & -0.406&   -0.453     \\  
 $D^+_s(c\bar{s})$ &   0.205&   0.216    \\
 $B^+(u\bar{b})$   &   0.490&  0.536     \\ 
 $B^0(d\bar{b})$   &  -0.242&  -0.266    \\
 $B^0_s(s\bar{b})$ &  -0.207&  -0.214    \\    \hline
\end{tabular}
\label{stability charge}
\end{table}

We have also checked the sensitivity of charge radii for different cut-off ($r_0$) values. The results are presented in Table 4.

\begin{table}[!h]\footnotesize
\label{sensitivity}
\caption{ The sensitivity of mean square charge radii of $D$ and $B$ mesons with different $r_0$ values. }
\begin{center}
\begin{tabular}{lllllllll}
\hline
                               Meson              &  \multicolumn{4}{c}{$\langle r^2 \rangle$ ($Fermi^2$)}   \\ \hline
            & $r_0= 0.689 Fermi$      & $r_0= 0.788 Fermi$ & $r_0= 1 Fermi$ & $r_0= 1.379 Fermi$ \\ \hline
 $D^+(c\bar{d}$)  &0.166 & 0.200&    0.260    &  0.307   \\ 
 $D^0(c\bar{u})$ &-0.289  & -0.349&   -0.453    & -0.535  \\  
 $D^+_s(c\bar{s})$&0.152 &   0.180&   0.216    &  0.305   \\
 $B^+(u\bar{b})$ &0.337  &   0.410&  0.536    &  0.617   \\ 
 $B^0(d\bar{b})$  &0.167 &  -0.204&  -0.266    &  -0.306 \\
 $B^0_s(s\bar{b})$&-0.167 &  -0.180&  -0.214    &  -0.287  \\    \hline

\end{tabular}
\end{center}
\end{table}

From Table 4, it is seen that the higher value of the cut-off $(r_0$) raises the charge radii of the mesons. It is clear that our results for charge radii of mesons agree well with those of ref.\cite{hwang} when the upper cut-off $r_0$ will be between 0.689 and 0.788 $Fermi$. Further it is to mention that the value of $r_0$ could not be less than that of $r^P$.

\section{Conclusion}

 %In ref. \cite{nsbijp}, the model did not show good results for the charge radii of $B$ mesons, though the model was quite successful in the prediction of results for light mesons and heavy flavored $D$ mesons. However the confinement parameter `$b$' was taken to be zero due to perturbative constraints. With the improved version of the present work, we overcome such limitation by incorporating the value of confinement parameter to be $b=0.183 GeV^2$ and also by introducing a cut-off parameter $r^P$. \\
We have studied the charge radii of various heavy flavored mesons $D^+(c\bar{d}$), $D^0(c\bar{u})$, $D^+_s(c\bar{s})$, $B^+(u\bar{b})$, $B^0(d\bar{b})$, $B^0_s(s\bar{b})$ in an improved version of a specific potential model\cite{pdas} in which the short range and long range effect of the Cornell potential is expected to be enhanced in the perturbative procedure.\\

The scale parameter `$r^P$' in the model is not an arbitrary parameter rather it depends on the strong coupling constant $\alpha_s$ and confinrmrnt parameter $b$. Again, we have used the same input parameters as is used in ref. \cite{my1, my2, my3}.\\

The charge radii of various heavy flavored mesons have not been measured experimentally yet. Our predicted results for charge radii of $D^+(c\bar{d})$ and $D^0(c\bar{u})$ mesons show good agreement to model of ref. \cite{lombard}. From Table 4, it is clear that adjusting the upper cut-off $r_0$ to $ \sim 0.788 Fermi$, our results agree with those of ref. \cite{hwang}. In the present work we tried to explain the physical significance of negative charge radii of neutral mesons. The graphs of Fig. \ref{fig2} show the variation of form factor $F(Q^2)$ with $Q^2$ for neutral mesons which indicate that the slope of the form factor is positive for $Q^2$ nearly upto $1 GeV^2$.\\ %After $Q^2 \approx 1 GeV^2$, the slope becomes negative, this might be a significance of neutral meson oscillation; $B^0-\bar{B^0}$ , $D^0-\bar{D^0}$ oscillation \cite{Boscillation, Doscillation}.\\

Inspite of its success in various stage, the model has a limitation that the relativistic effect of the quarks involved is not incorporated directly but by a modification factor $(\frac{r}{a_0})^{-\epsilon}$ in a free Dirac way, but there should also have some other significant dynamical effects. Further study needed to take into account such limitation.
%So this can be considered as one of the limitations of the present version of the work. Further though the model is found to be successful compared to the previous work, it has some limitations too. One of which is the full relativity could not be considered for heavy-light meson systems. Though the relativistic effect is introduced through the Dirac modification factor
\section*{Acknowledgement}
\textit{One of the authors (T Das) acknowledges the support of University Grants Commission in terms of fellowship under BSR scheme to pursue research work at Gauhati University, Department of Physics.}

\appendix
\numberwithin{equation}{section}
%\addappheadtotoc
%\appendixpage
\appendixpageoff
\section{Appendix}
\subsection*{With Coulomb parent linear parturbed wave function (9):}
\begin{equation}
F(Q^2)\mid _I=\sum_{i=1}^2 \frac{e_i}{Q_i}\int_0^{r^P} 4\pi r \mid\psi(r)\mid^2 sin(Q_ir)dr
\end{equation}

Using equation (9) in equation (A.1) and integrating over $r$,
\begin{equation}\footnotesize
\begin{split}
F(Q^2)\mid _I= N^2 \sum_{i=1}^2 \frac{e_i}{Q_i} \biggl[ \frac{2^{2\epsilon}}{a_0}(\gamma(2-2\epsilon, r^P))sin((2-2\epsilon).\theta_i){(1+\frac{a_0^2 Q_i^2}{4})^{\epsilon-1}} \\ 
+ \frac{a_0^5}{2^{6-2\epsilon}} \mu^2 b^2 (\gamma(6-2\epsilon, r^P)) sin((6-2\epsilon).\theta_i){(1+\frac{a_0^2 Q_i^2}{4})^{\epsilon-3}} \\ 
-\frac{a_0^2}{2^{2-2\epsilon}} \mu b (\gamma(4-2\epsilon, r^P))sin((4-2\epsilon).\theta_i)(1+\frac{a_0^2 Q_i^2}{4})^{\epsilon-2}\biggr]
\end{split}
\end{equation}

where 
\begin{equation}
\theta_i=sin^{-1}\left[ \frac{Q_i}{\left( Q_i^2+\frac{4}{a_0^2}\right)^{\frac{1}{2}} }\right] ,
\end{equation}

where only the first term of the following series is considered

\begin{equation}
sin^{-1}(x)\thickapprox x+\frac{x^3}{6}+\frac{3x^5}{40},
\end{equation}

with
\begin{equation}
x=\frac{Q_i}{\left( Q_i^2+\frac{4}{a_0^2}\right)^{\frac{1}{2}} }
\end{equation}

which is true for very low $Q^2$.\\

We split the $sine$ function of equation (A.2) using
\begin{equation}
siny=y- \frac{y^3}{3!}+ \frac{y^5}{5!}.
\end{equation}

Now equation (A.2) becomes
\begin{equation}\footnotesize
\begin{split}
F(Q^2)\mid _I= N^2 \sum_{i=1}^2 \frac{e_i}{Q_i} \biggl[ \frac{2^{2\epsilon}}{a_0}(\gamma(2-2\epsilon, r^P))\left((2-2\epsilon)\theta_i-\frac{(2-2\epsilon)^3}{3!}\theta_i^3+\frac{(2-2\epsilon)^5}{5!}\theta_i^5\right){(1+\frac{a_0^2 Q_i^2}{4})^{\epsilon-1}} \\ 
+ \frac{a_0^5}{2^{6-2\epsilon}} \mu^2 b^2 (\gamma(6-2\epsilon, r^P)) \left((6-2\epsilon)\theta_i-\frac{(6-2\epsilon)^3}{3!}\theta_i^3+\frac{(6-2\epsilon)^5}{5!}\theta_i^5\right){(1+\frac{a_0^2 Q_i^2}{4})^{\epsilon-3}} \\ 
-\frac{a_0^2}{2^{2-2\epsilon}} \mu b (\gamma(4-2\epsilon, r^P))\left((4-2\epsilon)\theta_i-\frac{(4-2\epsilon)^3}{3!}\theta_i^3+\frac{(4-2\epsilon)^5}{5!}\theta_i^5\right)(1+\frac{a_0^2 Q_i^2}{4})^{\epsilon-2}\biggr].
\end{split}
\end{equation}

Using (A.3) and (A.5) in the above equation,

\begin{equation}\footnotesize
\begin{split}
F(Q^2)\mid _I= N^2 \sum_{i=1}^2 e_i \biggl[ \frac{2^{2\epsilon}}{a_0}(\gamma(2-2\epsilon, r^P))\biggl((2-2\epsilon)X_i-
\frac{(2-2\epsilon)^3}{3!}Q_i^2 X_i^3+\frac{(2-2\epsilon)^5}{5!}Q_i^4X_i^5\biggr){(1+\frac{a_0^2 Q_i^2}{4})^{\epsilon-1}} \\ 
+ \frac{a_0^5}{2^{6-2\epsilon}} \mu^2 b^2 (\gamma(6-2\epsilon, r^P)) \biggl((6-2\epsilon)X_i
-\frac{(6-2\epsilon)^3}{3!}Q_i^2X_i^3+\frac{(6-2\epsilon)^5}{5!}Q_i^4X_i^5\biggr){(1+\frac{a_0^2 Q_i^2}{4})^{\epsilon-3}} \\ 
-\frac{a_0^2}{2^{2-2\epsilon}} \mu b (\gamma(4-2\epsilon, r^P))\biggl((4-2\epsilon)X_i
-\frac{(4-2\epsilon)^3}{3!}Q_i^2X_i^3+\frac{(4-2\epsilon)^5}{5!}Q_i^4X_i^5\biggr)(1+\frac{a_0^2 Q_i^2}{4})^{\epsilon-2}
\biggr]
\end{split}
\end{equation}

where
\begin{equation}
X_i=\left( Q_i^2+\frac{4}{a_0^2}\right)^{-\frac{1}{2}}.
\end{equation}

At low $Q^2$ limit, equation (A.8) reduces to equation (34).

\begin{equation}\footnotesize
\begin{split}
F(Q^2)\mid _I= N^2 \sum_{i=1}^2 e_i \biggl[ \frac{1}{2^{1-2\epsilon}}(\gamma(2-2\epsilon, r^P))(2-2\epsilon)(1+\frac{a_0^2 Q_i^2}{4})^{\epsilon-\frac{3}{2}} \\ 
-\frac{\mu b a_0^3}{2^{3-2\epsilon}}  (\gamma(4-2\epsilon, r^P)) (4-2\epsilon)(1+\frac{a_0^2 Q_i^2}{4})^{\epsilon-\frac{5}{2}} \\ 
+\frac{\mu^2 b^2 a_0^6}{2^{7-2\epsilon}}  (\gamma(6-2\epsilon, r^P))(6-2\epsilon)(1+\frac{a_0^2 Q_i^2}{4})^{\epsilon-\frac{7}{2}}
\biggr].
\end{split}
\end{equation}

Without relativistic effect ($\epsilon=0$) the above equation reduces to

\begin{equation}\footnotesize
\begin{split}
F(Q^2)\mid _I= N^2 \sum_{i=1}^2 e_i \biggl[\gamma(2, r^P)(1+\frac{a_0^2 Q_i^2}{4})^{-\frac{3}{2}} \\ 
-\frac{\mu b a_0^3}{2} \gamma(4, r^P)(1+\frac{a_0^2 Q_i^2}{4})^{-\frac{5}{2}} \\ 
+\frac{3\mu^2 b^2 a_0^6}{2^6} \gamma(6, r^P)(1+\frac{a_0^2 Q_i^2}{4})^{-\frac{7}{2}}
\biggr].
\end{split}
\end{equation}
\section{Appendix}
\subsection*{With linear parent Coulomb perturbed wave function (14) considering Airy function upto $O(r^1)$:}
%%%%%%%%%%%%%%

\begin{equation}
F(Q^2)\mid _{II}=\sum_{i=1}^2 \frac{e_i}{Q_i}\int_{r^P}^{r_0} 4\pi r \mid\psi(r)\mid^2 sin(Q_ir)dr
\end{equation}
Similarly, at low $Q^2$ limit equation (B.1) gives
\begin{equation}\footnotesize
\begin{split}
F(Q^2)\mid _{II}= 4 \pi N^{\prime 2} a_0^{2 \epsilon} \sum_{i=1}^2 e_i \biggl[ \frac{(a_1-b_1\rho_0)^2 (\gamma(-2\epsilon,r_0)-\gamma(-2\epsilon,r^P)) sin((-2\epsilon)\phi_i)}{(Q_i^2)^{\frac{1-2\epsilon}{2}}}\\
+ \frac{(b_1\rho_1)^2 (\gamma(2-2\epsilon,r_0)-\gamma(2-2\epsilon,r^P)) sin((2-2\epsilon)\phi_i)}{(Q_i^2)^{\frac{3-2\epsilon}{2}}}\\
-\frac{2b_1 \rho_1(a_1-b_1\rho_0) (\gamma(1-2\epsilon,r_0)-\gamma(1-2\epsilon,r^P)) sin((1-2\epsilon)\phi_i)}{(Q_i^2)^{\frac{2-2\epsilon}{2}}}\\
-\frac{16}{3}\frac{(a_1-b_1\rho_0)^2 \mu \alpha_s (\gamma(3-2\epsilon,r_0)-\gamma(3-2\epsilon,r^P)) sin((3-2\epsilon)\phi_i)}{(\rho_1k)^2 (Q_i^2)^{\frac{4-2\epsilon}{2}}}\\
+\frac{16}{3}\frac{(a_1-b_1\rho_0)^2 \mu \alpha_s 2 \sqrt{2\rho_1}(\gamma(4-2\epsilon,r_0)-\gamma(4-2\epsilon,r^P)) sin((4-2\epsilon)\phi_i)}{(\rho_1k)^3 (Q_i^2)^{\frac{5-2\epsilon}{2}}}\\
-\frac{16}{3}\frac{(b_1\rho_1)^2 \mu \alpha_s (\gamma(5-2\epsilon,r_0)-\gamma(5-2\epsilon,r^P)) sin((5-2\epsilon)\phi_i)}{(\rho_1k)^2 (Q_i^2)^{\frac{6-2\epsilon}{2}}}\\
+\frac{16}{3}\frac{(b_1\rho_1)^2 \mu \alpha_s 2 \sqrt{2\rho_1}(\gamma(6-2\epsilon,r_0)-\gamma(6-2\epsilon,r^P)) sin((6-2\epsilon)\phi_i)}{(\rho_1k)^3 (Q_i^2)^{\frac{7-2\epsilon}{2}}}\\
+\frac{32}{3}\frac{b_1 \rho_1 (a_1-b_1\rho_0) \mu \alpha_s (\gamma(4-2\epsilon,r_0)-\gamma(4-2\epsilon,r^P)) sin((4-2\epsilon)\phi_i)}{(\rho_1k)^2 (Q_i^2)^{\frac{5-2\epsilon}{2}}}\\
-\frac{32}{3}\frac{b_1 \rho_1 (a_1-b_1\rho_0) 2\sqrt{2\rho_1}\mu \alpha_s (\gamma(5-2\epsilon,r_0)-\gamma(5-2\epsilon,r^P)) sin((5-2\epsilon)\phi_i)}{(\rho_1k)^3 (Q_i^2)^{\frac{6-2\epsilon}{2}}}\\
-\frac{8}{3}\frac{(a_1-b_1\rho_0)^2 \mu \alpha_s (\gamma(4-2\epsilon,r_0)-\gamma(4-2\epsilon,r^P)) sin((4-2\epsilon)\phi_i)}{(\rho_1k)^2 (Q_i^2)^{\frac{5-2\epsilon}{2}}}\\
+\frac{8}{3}\frac{(a_1-b_1\rho_0)^2 \mu \alpha_s 2\sqrt{2\rho_1}(\gamma(5-2\epsilon,r_0)-\gamma(5-2\epsilon,r^P)) sin((5-2\epsilon)\phi_i)}{(\rho_1k)^3 (Q_i^2)^{\frac{6-2\epsilon}{2}}}\\
+(\frac{8\mu \alpha_s}{3})^2\frac{(b_1 \rho_1)^2 (\gamma(8-2\epsilon,r_0)-\gamma(8-2\epsilon,r^P)) sin((8-2\epsilon)\phi_i)}{(\rho_1k)^4 (Q_i^2)^{\frac{9-2\epsilon}{2}}}\\
+(\frac{8\mu \alpha_s}{3})^2\frac{8\rho_1(b_1 \rho_1)^2 (\gamma(10-2\epsilon,r_0)-\gamma(10-2\epsilon,r^P)) sin((10-2\epsilon)\phi_i)}{(\rho_1k)^6 (Q_i^2)^{\frac{11-2\epsilon}{2}}}\\
-(\frac{8\mu \alpha_s}{3})^2\frac{(b_1 \rho_1)^2 4\sqrt{2\rho_1}(\gamma(9-2\epsilon,r_0)-\gamma(9-2\epsilon,r^P)) sin((9-2\epsilon)\phi_i)}{(\rho_1k)^5 (Q_i^2)^{\frac{10-2\epsilon}{2}}}\\
-(\frac{8\mu \alpha_s}{3})^2\frac{2 b_1 \rho_1 (a_1-b_1\rho_0)(\gamma(7-2\epsilon,r_0)-\gamma(7-2\epsilon,r^P)) sin((7-2\epsilon)\phi_i)}{(\rho_1k)^4 (Q_i^2)^{\frac{8-2\epsilon}{2}}}\\
-(\frac{8\mu \alpha_s}{3})^2\frac{2 b_1 \rho_1 (a_1-b_1\rho_0)8\rho_1(\gamma(9-2\epsilon,r_0)-\gamma(9-2\epsilon,r^P)) sin((9-2\epsilon)\phi_i)}{(\rho_1k)^6 (Q_i^2)^{\frac{10-2\epsilon}{2}}}\\
+(\frac{8\mu \alpha_s}{3})^2\frac{4 b_1 \rho_1 (a_1-b_1\rho_0)2\sqrt{2\rho_1}(\gamma(8-2\epsilon,r_0)-\gamma(8-2\epsilon,r^P)) sin((8-2\epsilon)\phi_i)}{(\rho_1k)^5 (Q_i^2)^{\frac{9-2\epsilon}{2}}}\biggr],
\end{split}
\end{equation}

where 
\begin{center}

$\phi_i=sin^{-1}(1)$,\\

$a_1=0.355028$,\\

$b_1=0.258819$,\\

$b=0.183 GeV^2$,\\

$\rho_0=-\frac{3}{4} 3^{\frac{1}{3}} \pi^{\frac{2}{3}}$,\\

$\rho_1=0.715309\mu^{\frac{1}{3}}$\\

and $k=1.33586\mu^{\frac{1}{3}}.$\\

\end{center}

Putting above values in equation (B.2) and using approximations (A.4) and (A.6), the equation (B.2) reduces to equation (B.3)\\

\begin{equation}\footnotesize
\begin{split}
F(Q^2)\mid _{II}= 4 \pi N^{\prime 2} a_0^{2 \epsilon} \sum_{i=1}^2 e_i \biggl[0.913 (\gamma(-2\epsilon,r_0)-\gamma(-2\epsilon,r^P))(-2\epsilon)\frac{1}{(Q_i^2)^{\frac{1-2\epsilon}{2}}}\\
-0.353 \mu^{\frac{1}{3}}(\gamma(1-2\epsilon,r_0)-\gamma(1-2\epsilon,r^P)) (1-2\epsilon)\frac{1}{(Q_i^2)^{\frac{2-2\epsilon}{2}}}\\
+0.0342\mu^{\frac{2}{3}}(\gamma(2-2\epsilon,r_0)-\gamma(2-2\epsilon,r^P)) (2-2\epsilon)\frac{1}{(Q_i^2)^{\frac{3-2\epsilon}{2}}}\\
-5.33 \mu\alpha_s(\gamma(3-2\epsilon,r_0)-\gamma(3-2\epsilon,r^P)) (3-2\epsilon)\frac{1}{(Q_i^2)^{\frac{4-2\epsilon}{2}}}\\
+(13.35\mu^{\frac{7}{6}}+2.06\mu^{\frac{4}{3}}-2.66\mu)\alpha_s (\gamma(4-2\epsilon,r_0)-\gamma(4-2\epsilon,r^P)) (4-2\epsilon)\frac{1}{(Q_i^2)^{\frac{5-2\epsilon}{2}}}\\
+(6.675 \mu^{\frac{7}{6}}-5.17 \mu^{\frac{3}{2}}-0.2 \mu^{\frac{5}{3}}) \alpha_s(\gamma(5-2\epsilon,r_0)-\gamma(5-2\epsilon,r^P)) (5-2\epsilon)\frac{1}{(Q_i^2)^{\frac{6-2\epsilon}{2}}}\\
+0.501 \mu^{\frac{11}{6}}\alpha_s(\gamma(6-2\epsilon,r_0)-\gamma(6-2\epsilon,r^P)) (6-2\epsilon)\frac{1}{(Q_i^2)^{\frac{7-2\epsilon}{2}}}\\
-3.017 \mu^{\frac{7}{3}}\alpha_s^2 (\gamma(7-2\epsilon,r_0)-\gamma(7-2\epsilon,r^P)) (7-2\epsilon)\frac{1}{(Q_i^2)^{\frac{8-2\epsilon}{2}}}\\
+(0.292 \mu^{\frac{8}{3}}+15.1\mu^{\frac{5}{2}})\alpha_s^2 (\gamma(8-2\epsilon,r_0)-\gamma(8-2\epsilon,r^P)) (8-2\epsilon)\frac{1}{(Q_i^2)^{\frac{9-2\epsilon}{2}}}\\
-(1.463 \mu^{\frac{17}{6}}+18.91 \mu^{\frac{8}{3}})\alpha_s^2 (\gamma(9-2\epsilon,r_0)-\gamma(9-2\epsilon,r^P)) (9-2\epsilon)\frac{1}{(Q_i^2)^{\frac{10-2\epsilon}{2}}}\\
+1.83 \mu^3 \alpha_s^2 (\gamma(10-2\epsilon,r_0)-\gamma(10-2\epsilon,r^P)) (10-2\epsilon)\frac{1}{(Q_i^2)^{\frac{11-2\epsilon}{2}}}
\biggr].
\end{split}
\end{equation}

\begin{equation}\footnotesize
\begin{split}
F(Q^2)\mid _{II}= 4 \pi N^{\prime 2} a_0^{2 \epsilon} \sum_{i=1}^2 e_i \biggl[F_1\frac{1}{(Q_i^2)^{\frac{1-2\epsilon}{2}}}
+F_2\frac{1}{(Q_i^2)^{\frac{2-2\epsilon}{2}}}
+F_3\frac{1}{(Q_i^2)^{\frac{3-2\epsilon}{2}}}
+F_4\frac{1}{(Q_i^2)^{\frac{4-2\epsilon}{2}}}\\
+F_5\frac{1}{(Q_i^2)^{\frac{5-2\epsilon}{2}}}
+F_6\frac{1}{(Q_i^2)^{\frac{6-2\epsilon}{2}}}
+F_7\frac{1}{(Q_i^2)^{\frac{7-2\epsilon}{2}}}
+F_8\frac{1}{(Q_i^2)^{\frac{8-2\epsilon}{2}}}\\
+F_9\frac{1}{(Q_i^2)^{\frac{9-2\epsilon}{2}}}
+F_{10}\frac{1}{(Q_i^2)^{\frac{10-2\epsilon}{2}}}
+F_{11}\frac{1}{(Q_i^2)^{\frac{11-2\epsilon}{2}}}
\biggr],
\end{split}
\end{equation}

where
\begin{equation}\footnotesize
\begin{split}
F_1=0.913 (\gamma(-2\epsilon,r_0)-\gamma(-2\epsilon,r^P))(-2\epsilon)\\
F_2=-0.353 \mu^{\frac{1}{3}}(\gamma(1-2\epsilon,r_0)-\gamma(1-2\epsilon,r^P)) (1-2\epsilon)\\
F_3=0.0342\mu^{\frac{2}{3}}(\gamma(2-2\epsilon,r_0)-\gamma(2-2\epsilon,r^P)) (2-2\epsilon)\\
F_4=-5.33 \mu\alpha_s(\gamma(3-2\epsilon,r_0)-\gamma(3-2\epsilon,r^P)) (3-2\epsilon)\\
F_5=(13.35\mu^{\frac{7}{6}}+2.06\mu^{\frac{4}{3}}-2.66\mu)\alpha_s (\gamma(4-2\epsilon,r_0)-\gamma(4-2\epsilon,r^P)) (4-2\epsilon)\\
F_6=(6.675 \mu^{\frac{7}{6}}-5.17 \mu^{\frac{3}{2}}-0.2 \mu^{\frac{5}{3}}) \alpha_s(\gamma(5-2\epsilon,r_0)-\gamma(5-2\epsilon,r^P)) (5-2\epsilon)\\
F_7=0.501 \mu^{\frac{11}{6}}\alpha_s(\gamma(6-2\epsilon,r_0)-\gamma(6-2\epsilon,r^P)) (6-2\epsilon)\\
F_8=-3.017 \mu^{\frac{7}{3}}\alpha_s^2 (\gamma(7-2\epsilon,r_0)-\gamma(7-2\epsilon,r^P)) (7-2\epsilon)\\
F_9=(0.292 \mu^{\frac{8}{3}}+15.1\mu^{\frac{5}{2}})\alpha_s^2 (\gamma(8-2\epsilon,r_0)-\gamma(8-2\epsilon,r^P)) (8-2\epsilon)\\
F_{10}=-(1.463 \mu^{\frac{17}{6}}+18.91 \mu^{\frac{8}{3}})\alpha_s^2 (\gamma(9-2\epsilon,r_0)-\gamma(9-2\epsilon,r^P)) (9-2\epsilon)\\
F_{11}=1.83 \mu^3 \alpha_s^2 (\gamma(10-2\epsilon,r_0)-\gamma(10-2\epsilon,r^P)) (10-2\epsilon).
\end{split}
\end{equation}

We can express equation (B.4) as
\begin{equation}
\begin{split}
F(Q^2)\mid _{II}= 4 \pi N^{\prime 2} a_0^{2 \epsilon} \sum_{i=1}^2 e_i \biggl[
\sum_{k=1}^{11} F_k\frac{1}{(Q_i^2)^{\frac{k-2\epsilon}{2}}}
\biggr].
\end{split}
\end{equation}

Without relativistic effect ($\epsilon=0$) the above equation reduces to
\begin{equation}
F(Q^2)\mid _{II}= 4 \pi N^{\prime 2} \sum_{i=1}^2 e_i \biggl[
\sum_{k=2}^{11} F^\prime_k\frac{1}{(Q_i^2)^{\frac{k}{2}}}
\biggr],
\end{equation}

where
\begin{equation}\footnotesize
\begin{split}
F^\prime_2=-0.353 \mu^{\frac{1}{3}}(\gamma(1,r_0)-\gamma(1,r^P)) \\
F^\prime_3=2\times0.0342\mu^{\frac{2}{3}}(\gamma(2,r_0)-\gamma(2,r^P)) \\
F^\prime_4=-3\times5.33 \mu\alpha_s(\gamma(3,r_0)-\gamma(3,r^P)) \\
F^\prime_5=4\times(13.35\mu^{\frac{7}{6}}+2.06\mu^{\frac{4}{3}}-2.66\mu)\alpha_s (\gamma(4,r_0)-\gamma(4,r^P))\\
F^\prime_6=5\times(6.675 \mu^{\frac{7}{6}}-5.17 \mu^{\frac{3}{2}}-0.2 \mu^{\frac{5}{3}}) \alpha_s(\gamma(5,r_0)-\gamma(5,r^P)) \\
F^\prime_7=6\times0.501 \mu^{\frac{11}{6}}\alpha_s(\gamma(6,r_0)-\gamma(6,r^P))\\
F^\prime_8=-7\times3.017 \mu^{\frac{7}{3}}\alpha_s^2 (\gamma(7,r_0)-\gamma(7,r^P)) \\
F^\prime_9=8\times(0.292 \mu^{\frac{8}{3}}+15.1\mu^{\frac{5}{2}})\alpha_s^2 (\gamma(8,r_0)-\gamma(8,r^P)) \\
F^\prime_{10}=-9\times(1.463 \mu^{\frac{17}{6}}+18.91 \mu^{\frac{8}{3}})\alpha_s^2 (\gamma(9,r_0)-\gamma(9,r^P))\\
F^\prime_{11}=10\times1.83 \mu^3 \alpha_s^2 (\gamma(10,r_0)-\gamma(10,r^P)).
\end{split}
\end{equation}

In obtaining (A.8), (A.11), (B.6) and (B.7) we have also used the following integration
%\begin{equation}
%\int_0^x t^{s-1} e^{-t}dt= \gamma(s,x),
%\end{equation}

%\begin{equation}
%\int_u^v t^{s-1} e^{-t}dt= \gamma(s,v)-\gamma(s,u)
%\end{equation}

\begin{equation}
\int_0^\infty
x^{p-1} e^{-ax} sin(mx) dx= \frac{\Gamma(p) sin(p\theta)}{(a^2+m^2)^\frac{p}{2}}.
\end{equation}

The following form of Incomplete Gamma Function is used in obtaining (B.5) and (B.8)
\begin{equation}
\int_u^v t^{s-1} e^{-t}dt= \gamma(s,v)-\gamma(s,u).
\end{equation}

\end{document}